\documentclass[11pt]{amsart}
\usepackage{amsaddr} 
\usepackage{amsmath, amssymb, amsthm}
\usepackage{graphicx}
\usepackage{hyperref}
\usepackage{geometry}
\usepackage{enumitem} 
\usepackage{natbib}
\AtBeginDocument{
    \setlength{\abovedisplayskip}{15pt}
    \setlength{\belowdisplayskip}{15pt}
    \setlength{\abovedisplayshortskip}{10pt}
    \setlength{\belowdisplayshortskip}{10pt}
    \setlength{\jot}{8pt} % Spacing inside align environments
}

\newtheorem{theorem}{Theorem}

\newtheorem{example}{Example}
\newtheorem{remark}{Remark}

\title{Genealogical Reinforcement in Branching Processes}
\author{Alina Kochocki}
\address{Department of Physics, Loyola Marymount University, Los Angeles, CA 90045, USA}
\email{Alina.Kochocki@lmu.edu}
\date{\today}

\begin{document}

\begin{abstract}
Within many ecological populations, survival and reproduction exhibit a dependence on social interactions with relatives. As one example, resource excess is often shared according to genealogical proximity. When viewed as a branching system, an individual’s offspring distribution acquires an effective dependence on genealogical structure. In the specific case that a large, local family structure leads to a large set of offspring, this can be viewed as a form of genealogical reinforcement. To capture this mechanism, we introduce a branching model of finite-depth genealogical dependence. Given partial observations of this reproducing population and its genealogy, we would like to reject the hypothesis of independent branching and estimate the strength of dependence. We present a statistical framework for detection and estimation, combining analytical results with simulation-based validation. We then identify our branching model with a tree-valued multitype branching process on a finite type space, providing a rigorous characterization of the expected structural reinforcement. Finally, we discuss how these tree-valued interactions can be accommodated within a broader class of stochastic processes.
\end{abstract}
\maketitle

\section{Introduction}
Biological populations exhibit a dependence on their environment, a nonlinear coupling between the dynamics of the population and relevant resources. In species with complex social interactions, resource excess is often shared according to genealogical proximity. As individuals generally come from genealogical trees with substantial structural variation, this genealogical dependence introduces varied access to randomly distributed resources. An individual's expected resource coverage is directly dependent on the tree structure. 

Assuming a minimal resource threshold is required for survival, the expected number of offspring produced by an individual in a lifetime is closely tied to this resource expectation. If this continuous reproduction is instead viewed as a single, representative branching event, the mean number of offspring becomes a function of the genealogical tree. 

In these populations, evolution of the tree structure is effectively dependent on the existing tree structure \citep{thresh}. In the specific case of resource transfer prioritized by genealogical proximity, the influence of relatives on an individual's resource expectation decays with genealogical distance. A large family structure at short genealogical distances is most likely to beget a locally large family. We refer to this concept as genealogical reinforcement. The long-time behavior of the process results in a stable distribution of local tree types. 

In this work, we introduce the concept of genealogical dependence in a branching population from both a statistical and probabilistic perspective. We highlight that this system is not intended as a comprehensive social model. Instead, it isolates the dynamics of one mechanism present in many social-biological systems. 

\section{A Model of Branching with Finite-Depth Dependence}

We focus on branching in discrete time with dependence on the tree structure at finite depth \citep{athreya1972branching}. We explore a simple toy model for genealogical reinforcement with bounded numbers of offspring.  

Let $\mathcal{T}_{n}$ denote the genealogical tree of the population at generation $n$. Each vertex, $v$, corresponds to an individual, and edges represent parent-offspring relationships. For two individuals, $u$ and $v$, let $\text{MRCA}(u,v)$ denote their most recent common ancestor, let $k$ denote the generational distance between relatives of direct descent, and define the genealogical distance, 
\begin{equation}
    d(u,v) = k(v, \textrm{MRCA}(u,v)) + k(u, \textrm{MRCA}(u,v)).
\end{equation}
This is a standard path-length distance on a tree. When all individuals, $u, v,$ and $w$, are sampled from the same generation, this distance is ultrametric and satisfies the strong triangle inequality, 
\begin{equation}
d(u,v)\leq \, \max \left( d(u,w),d(w,v) \right).
\end{equation}

We assume individuals may have a number of offspring, $\xi(v) \in \{0,1,2 \}$. As the maximal offspring number, $K = 2$, is bounded, the local tree of depth $D$ has at most $2^{D}$ nodes. The entire historical genealogical tree can be represented by a tree of subtrees of depth $D$ on a finite type space, $\mathcal{K}_{D}$. 

Branching depends on a functional, $G(v, \mathcal{T}_{n})$, which captures the local genealogical configuration around an individual $v$. Let $\mathcal{L}_n$ denote the set of living individuals in generation $n$. To model genealogical reinforcement, we assume that the influence of a living relative decreases with genealogical distance. Specifically, we define:
\begin{equation}
    G(v, \mathcal{T}_{n}) = \sum_{\substack{u \in \mathcal{L}_n \\ u \neq v \\ d(u,v) \leq 2D} } e ^{- \theta  d(u,v)}.
\end{equation}
Here, $\theta \geq 0$ is a decay parameter of dependence. $D$ establishes the range of finite-depth dependence or truncation. $G$ is not normalized by this depth, ensuring the influence of close relatives does not decrease with an extended dependence. Small values of $\theta$ minimize the variation in dependence across the finite depth $D$. Large values of $\theta$ introduce a highly localized dependence within the subtree. 

\begin{figure}[htbp]
    \centering
    
    % --- Panel A ---
    \begin{minipage}{0.32\textwidth}
        \centering
        \includegraphics[width=\textwidth]{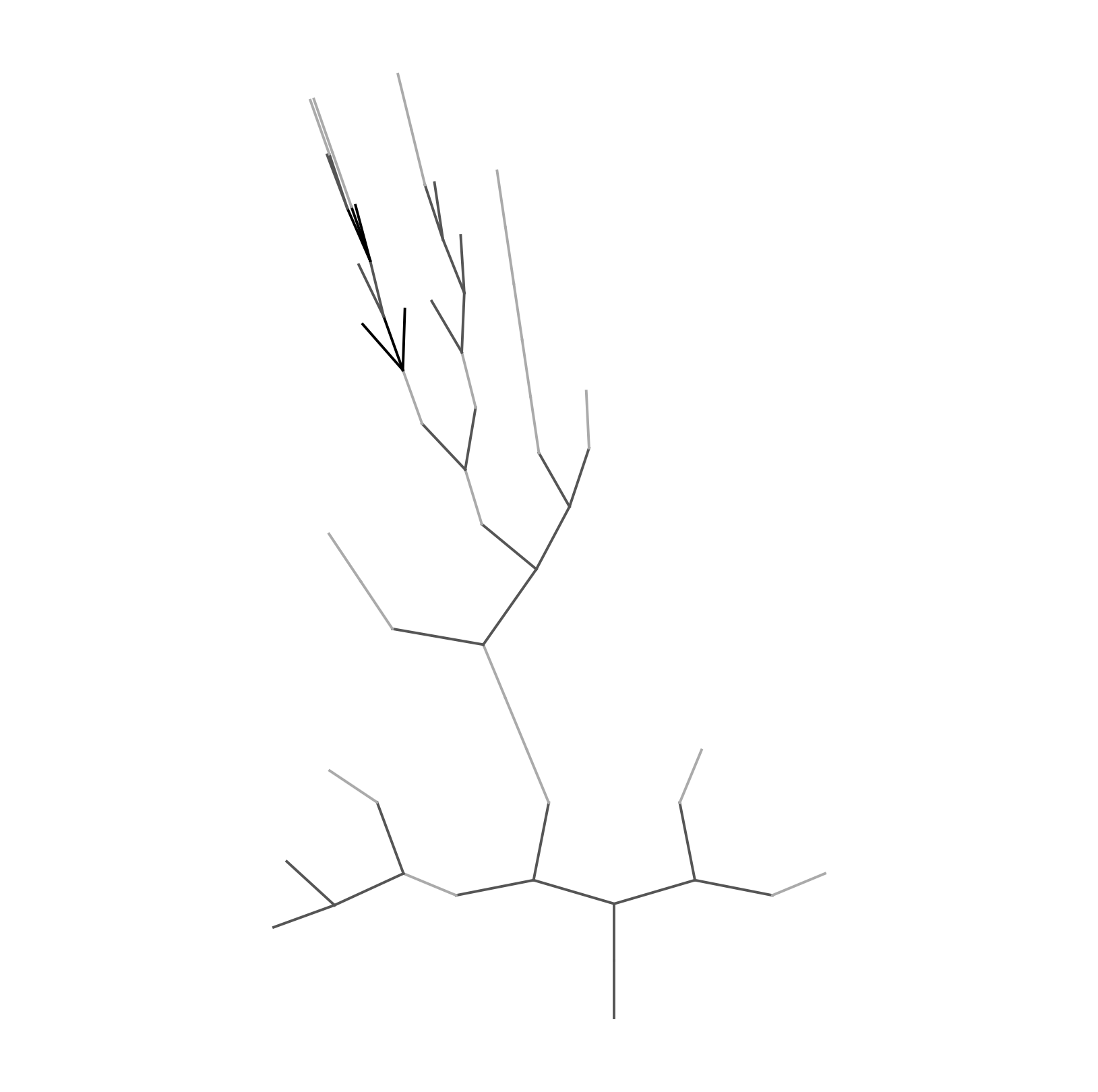}
        \\ \vspace{0.1cm} (a) Independent (\(\beta_1 = 0.0\))
    \end{minipage}\hfill
    % --- Panel B ---
    \begin{minipage}{0.32\textwidth}
        \centering
        \includegraphics[width=\textwidth]{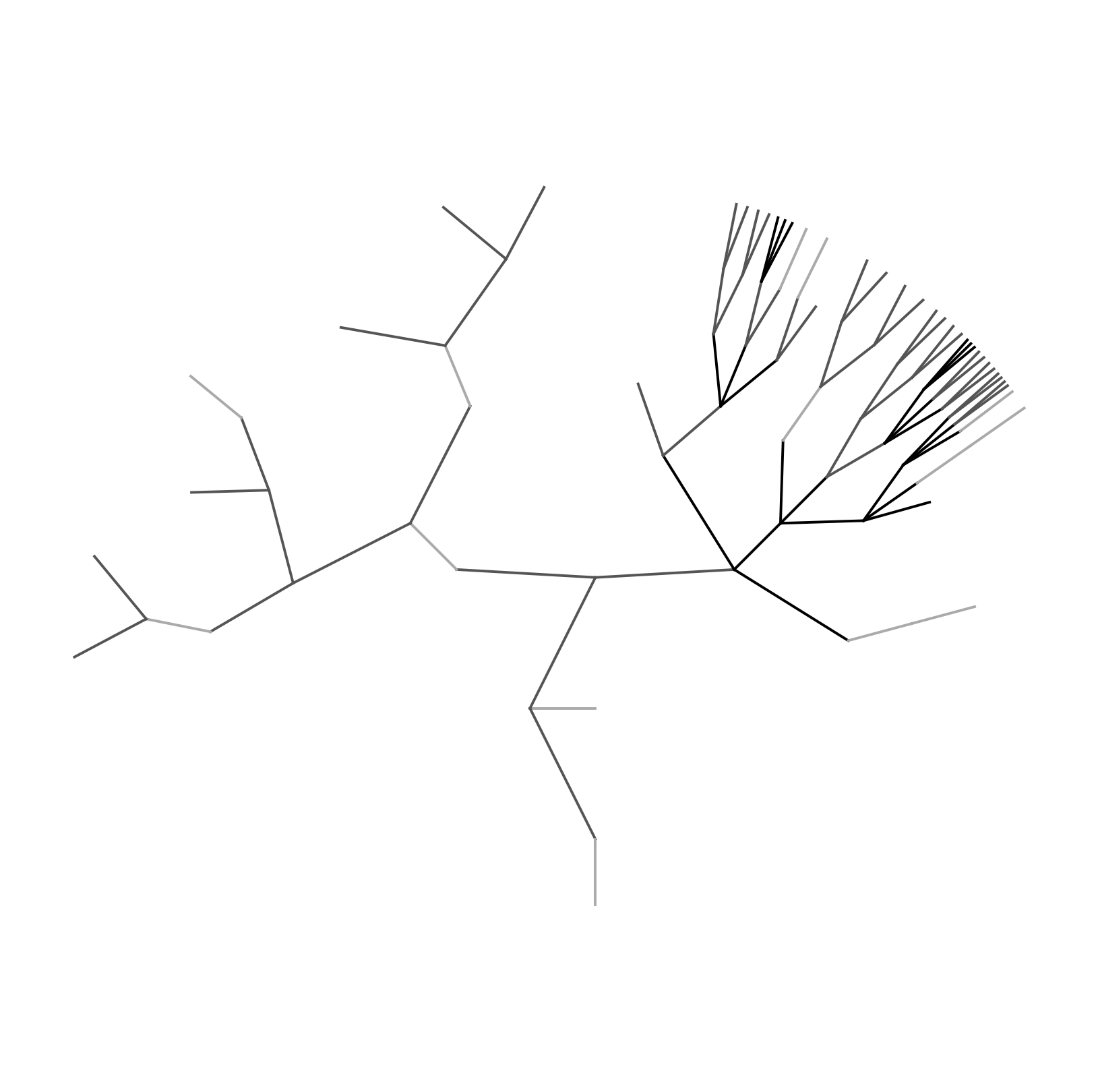}
        \\ \vspace{0.1cm} (b) Moderate (\(\beta_1 = 1.0\))
    \end{minipage}\hfill
    % --- Panel C ---
    \begin{minipage}{0.32\textwidth}
        \centering
        \includegraphics[width=\textwidth]{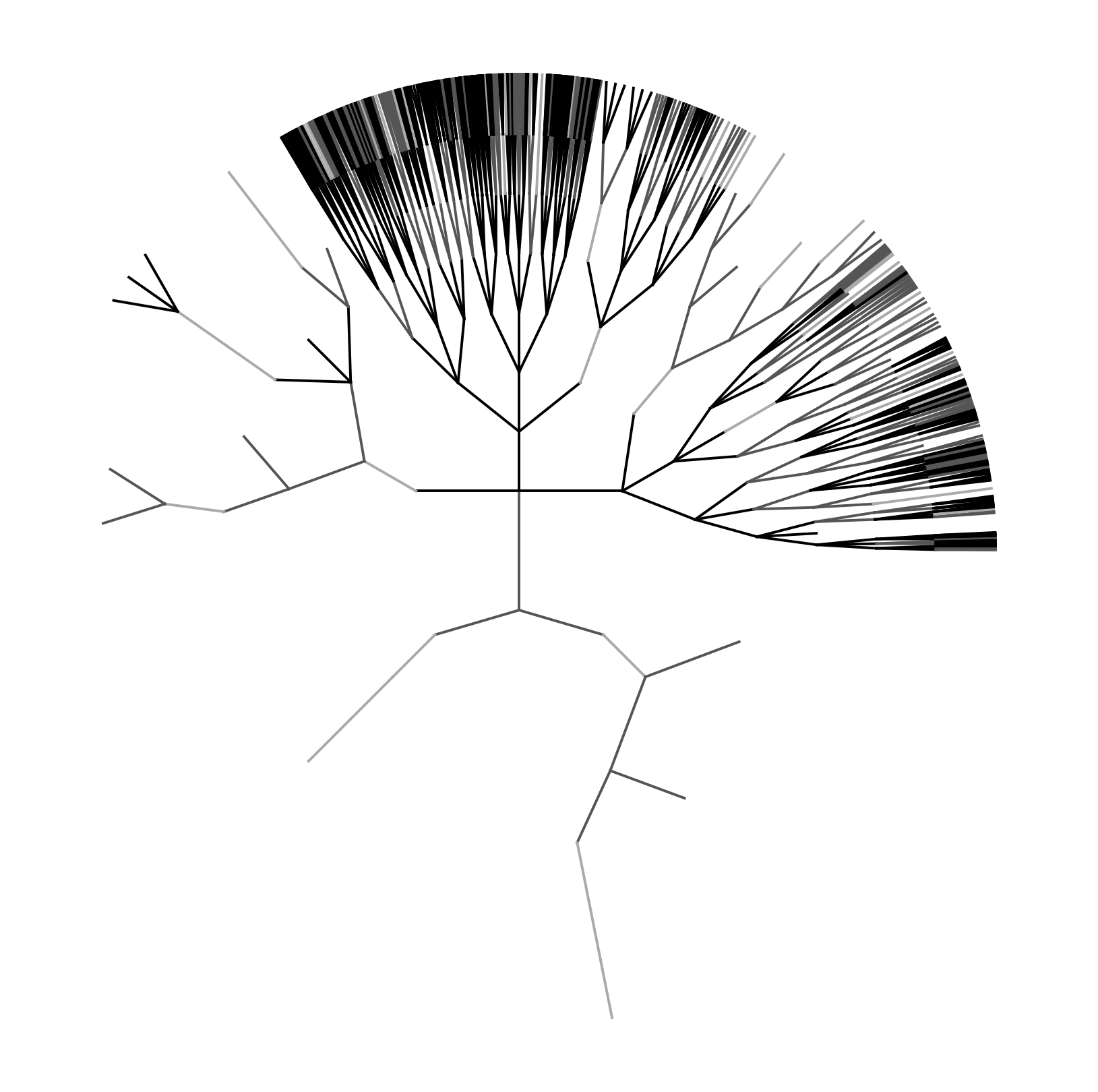}
        \\ \vspace{0.1cm} (c) Strong (\(\beta_1 = 1.5\))
    \end{minipage}
    
    \vspace{0.3cm}
    \caption{\textbf{Structural feedback in genealogical reinforcement.} Radial, simulated genealogical trees $(D=3, \, K=3, \, \theta = 0.5)$ under a subcritical baseline $(\beta_0 = -1.0)$. Edge color indicates the number of offspring (light gray = 1, black = 3). (a) Under independent branching, the lineage is sparse and randomly dispersed. (b) With moderate reinforcement, localized kin clusters begin to persist. (c) Under strong reinforcement, positive feedback drives runaway growth in specific families or subtrees (black clusters), while isolated lineages die out.}
    \label{fig:tree_panels}
\end{figure}

The functional $G$ enters the offspring distribution through the mean. To strictly enforce the bounded offspring assumption, we model the number of offspring as $\xi(v) \sim \textrm{Binomial}(K=2,p(v))$, where
\begin{equation}
    p(v) = \dfrac{1}{1 + \exp( -\beta_{0} - \beta_{1} G(v, \mathcal{T}_{n} ))} .
\end{equation}
Here, $\beta_{1} > 0$ controls the strength of genealogical reinforcement, and $\beta_{0}$ is a baseline reproduction rate. If $\beta_{1} = 0$, independent branching is recovered, similar to a traditional Galton-Watson process. The case of $\beta_{0} < 0$ represents a harsh baseline environment, where genealogical reinforcement is required to recover the population from extinction. We provide examples of simulated trees in Figure \ref{fig:tree_panels}. 

\begin{remark}[General Structural Reinforcement]
    Our model is motivated by the reinforcement of large family structures, where the maximally populated subtree dominates the long-time distribution of tree types. We note that models of genealogical reinforcement are not limited to this assumption. Offspring probabilities can be chosen to favor any other tree structure or distribution of types, allowing for general reinforcement on the subtree type space.
\end{remark}

If $\theta$ is small, the probability of offspring of each member of the subtree is effectively equivalent, and depends only on the size of this group. Large families tend to grow, regardless of structure. If $\theta$ is large, asymmetry of the subtree introduces an asymmetry in the offspring probability. 

In summary, our branching process is parameterized by the truncation depth of genealogical dependence, $D$, the maximal number of children, $K$, the decay of dependence, $\theta$, the strength of dependence, $\beta_{1}$, and a baseline reproduction rate, $\beta_{0}$. We treat $D$ and $K$ as fixed structural parameters to ensure a finite Markov state space. The continuous parameters $(\theta, \beta_{0}, \beta_{1})$ govern the biological dynamics, and subcritical, critical, or supercritical growth regimes. This local, tree-valued genealogical reinforcement contrasts with a recent body of work focused on the reinforcement of offspring probabilities sampled from the entire historical ancestry \citep{2023arXiv230602476B, 10.1214/24-AAP2144, bertoin2025reinforcedgaltonwatsonprocessesiii, bertoin2026solvablepopulationdynamicuniform}. 

\section{Statistical Detection Framework}

\subsection{Interaction Encoded Through the Offspring Distribution}

In the most general case, branching might depend on an individual's number of relatives of genealogical relationship type, $\tau$. In a continuous-time setting, multiple living generations may interact. We present a statistical framework to accommodate these generalized dependencies. For an individual $v$ at time $t$, let
\begin{equation}
    \mathbf{c}_{v} = (c_{\tau}(v))_{\tau \in \mathcal{T}}
\end{equation}
denote the vector of counts of relatives of each relationship type, $\tau \in \mathcal{T}$. The probability of reproduction may be a function of $\mathbf{c}_{v}$. To make inference tractable, we parameterize the log-odds of reproduction through a log-linear model with baseline, pairwise, and higher-order interactions:
\begin{equation}
    \operatorname{logit}(p(\mathbf{c}_{v})) = \alpha_{0} + \sum_{\tau} \alpha_{\tau} c_{\tau} + \sum_{\tau_{1} \leq \tau_{2}} \alpha_{\tau_{1}, \tau_{2}} c_{\tau_{1}} c_{\tau_{2}} + \cdots,
\end{equation}
where the coefficients $\alpha_{\tau}$ and $\alpha_{\tau_{1}, \tau_{2}}$ encode the strength of dependence on each relationship type and on interactions between types. These higher order terms reflect a joint dependence on the number of relatives of different relationship types. The null hypothesis of independent branching corresponds to $\alpha_{\tau} = 0$ for all $\tau$ (and similarly for all higher-order terms). 

The framework above is general in relationship type, $\tau$. For the model of genealogical reinforcement introduced in Section 2, we specialize to dependence on genealogical distance alone. Within a single generation, a given relationship type between living relatives is identified by their genealogical distance. The vector of counts in $\tau$ can be replaced by a vector of counts in genealogical distance. We identify $\alpha_{0} = \beta_{0}$ as the baseline reproduction rate. For relatives at distance $d$, the general coefficient $\alpha_{\tau} = \beta_{1} e^{-\theta d}$. Only these baseline and first-order interaction coefficients contribute. 

\subsection{Maximum Likelihood Estimation}

When the process is fully observed, detection of genealogical dependence proceeds through maximum likelihood estimation. While the general framework is parameterized by the baseline $\alpha_0$ and the coefficient vector $\vec{\alpha}$, our specific model of reinforcement constrains these parameters via the mapping $\alpha_0 = \beta_0$ and $\alpha_\tau = \beta_1 e^{-\theta d}$. The likelihood of our model is parameterized entirely by the scalar parameters $(\beta_{0}, \beta_{1}, \theta)$. 

Given the model and a sufficient initial condition, the likelihood of the fully observed trajectory factors over the branching events. Under our constrained reinforcement model, the likelihood is parameterized as,
\begin{equation}
    \mathcal{L}(\beta_0, \beta_1, \theta) = \prod_{v \in \mathcal{T}_n} \binom{2}{\xi(v)} p(v; \beta_0, \beta_1, \theta)^{\xi(v)} \big(1 - p(v; \beta_0, \beta_1, \theta)\big)^{2 - \xi(v)},
\end{equation}
where $\xi(v) \in \{0, 1, 2\}$ denotes the observed number of offspring of individual $v$, and $p(v)$ is the probability of reproduction defined in Section 2. As the offspring number is bounded, the likelihood is simply a product of Binomial probabilities over the observed tree.

In the finite-depth regime, the probability $p(v)$ depends only on the local tree structure of depth $D$ around the individual $v$. To constrain our three-parameter model $(\beta_0, \beta_1, \theta)$, a minimum of three observed branching events with sufficiently independent genealogy are required. To break the degeneracy between the baseline reproduction rate, $\beta_{0}$, reinforcement strength, $\beta_1$, and the decay rate, $\theta$, the observed data must exhibit sufficient structural variation. We formalize these structural conditions for identifiability in the following section.

\section{Identifiability Given Partial Data}

In our specific model of reinforcement, we constrain parameters $\beta_{0}$, $\beta_{1}$ and $\theta$ given the truncation depth of influence, $D$. Genealogical dependence is a simple monotonic function of distance, decaying exponentially as $e^{-\theta d}$. 

\begin{theorem}[Identifiability of the Reinforcement Model]
Assume the truncation depth satisfies $D \geq 2$. Suppose we observe the local subtrees and subsequent branching events of a sample of individuals. The parameters $\beta_0$, $\beta_1$, and $\theta$ are globally structurally identifiable from the likelihood $\mathcal{L}(\beta_0, \beta_1, \theta)$ if the sample falls into at least one of the following two cases:
\begin{enumerate}
    \item[(a)] One individual has $\mathbf{c}_v = \mathbf{0}$, and the other two individuals have linearly independent count vectors, $\mathbf{c}_v$, that contain at least two distinct, nonzero genealogical distances (e.g., \(d=2\) and \(d=4\)).
    \item[(b)] All three individuals have linearly independent count vectors, $\mathbf{c}_v$, and across the sample, at least three distinct, nonzero genealogical distances are observed.
\end{enumerate}
\end{theorem}

\begin{proof}[Proof Sketch]
We consider a non-trivial, multi-generational branching dependence. If $D = 0$, the model recovers independent branching, and $\beta_{1}$ and $\theta$ are not observable. The result is similar for $\beta_{1} = 0$. In the case of a truncation depth, $D = 1$, individuals depend only on their siblings with $\alpha_{1} = \beta_{1}e^{-\theta}$. The parameters are degenerate and unidentifiable. 

In the case $D \geq 2$, the log-odds of reproduction for an individual $v$ are a weighted sum over multiple distances. Let us denote the linear predictor by:
\begin{equation}
    \eta_v = \beta_0 + \beta_1 \sum_{d=1}^{2D} \mathbf{c}_v(d) e^{-\theta d}.
\end{equation}
where $\mathbf{c}_v(d)$ is the observed number of relatives at distance $d$. We consider constraint of this equation for our two cases. 

\textbf{Case (a):} The individual with $\mathbf{c}_v = \mathbf{0}$ has a linear predictor $\eta_v = \beta_0$. This directly anchors the baseline reproduction rate, decoupling it from the parameters of genealogical reinforcement. The remaining two individuals have linearly independent count vectors, providing two independent constraints. The condition that their vectors contain at least two distinct, nonzero distances ensures that the map \(\theta \mapsto (e^{-d_1\theta}, e^{-d_2\theta})\) is injective, decoupling $\theta$ from the scale parameter $\beta_1$.

\textbf{Case (b):} When all three count vectors are linearly independent, the system of equations for $(\beta_0, \beta_1)$ is uniquely solvable for any fixed $\theta$. The requirement that at least three distinct distances are observed provides sufficient variation to ensure that the likelihood for $\theta$ has a unique global maximum, preventing degenerate scaling with $\beta_1$. 

In both cases, the Fisher Information Matrix for $(\beta_0, \beta_1, \theta)$ is non-singular. This strict curvature guarantees \textit{local} structural identifiability. To establish \textit{global} identifiability, we note that the parameter $\theta$ is constrained to be positive (\(\theta > 0\)). The map \(\theta \mapsto e^{-\theta d}\) is strictly monotonic. This injectivity ensures a unique global maximum for our expected likelihood. The explicit calculation of the score functions and the non-singularity of the Fisher Information Matrix for these minimal samples are provided in Appendix A.
\end{proof}

\section{Rejection of Independent Branching}

\subsection{Likelihood Ratio Test}

We would like to establish whether genealogical dependence is present in a test population. We formalize this as a hypothesis test comparing the null model of independent branching to the alternative of our specific model of genealogical reinforcement.

The null hypothesis is given by $H_0: \beta_1 = 0$, which corresponds to independent branching. The alternative is $H_1: \beta_1 > 0$, where reproduction depends on the local genealogical structure. We construct the Likelihood Ratio Test (LRT) statistic, $-2 \ln \Lambda$:
\begin{equation}
    \Lambda = \left( \frac{\sup_{\beta_0} \mathcal{L}(\beta_0, \beta_1 = 0)}{\sup_{\beta_0, \beta_1 > 0, \theta \ge 0} \mathcal{L}(\beta_0, \beta_1, \theta)} \right).
\end{equation}

Standard asymptotic theory (Wilks' theorem) dictates that $-2 \ln \Lambda$ converges to a $\chi^2$ distribution \citep{Wilks:1938dza}. However, our hypothesis test violates the standard regularity conditions in two ways. First, the parameter $\beta_1$ lies on the boundary of the parameter space $(\beta_1 \ge 0)$. Second, under the null hypothesis, the decay parameter $\theta$ is unidentifiable, acting as a nuisance parameter that is only present under the alternative.

Given a model with a parameter on the boundary of the space, Chernoff's theorem dictates that the null distribution follows a mixture of a point mass at zero (denoted $\chi^2_0$) and a $\chi^2_1$ distribution, written as $\frac{1}{2}\chi^2_0 + \frac{1}{2}\chi^2_1$ \citep{Chernoff:1954eli}. However, as our model incorporates a nuisance parameter that is unidentifiable under the null hypothesis, we must instead directly simulate the null distribution of $\Lambda$. We simulate multiple independent branching populations $(\beta_1 = 0)$, compute $\Lambda$ for each, and empirically calibrate the rejection region. We use this calibrated LRT to characterize the statistical power of detecting genealogical reinforcement as a function of $\beta_{1}$, $\theta$, and tree depth, $D$. 

\subsection{Empirical Null Distribution}
\begin{figure}[th]
    \centering
    \includegraphics[width=0.8\textwidth]{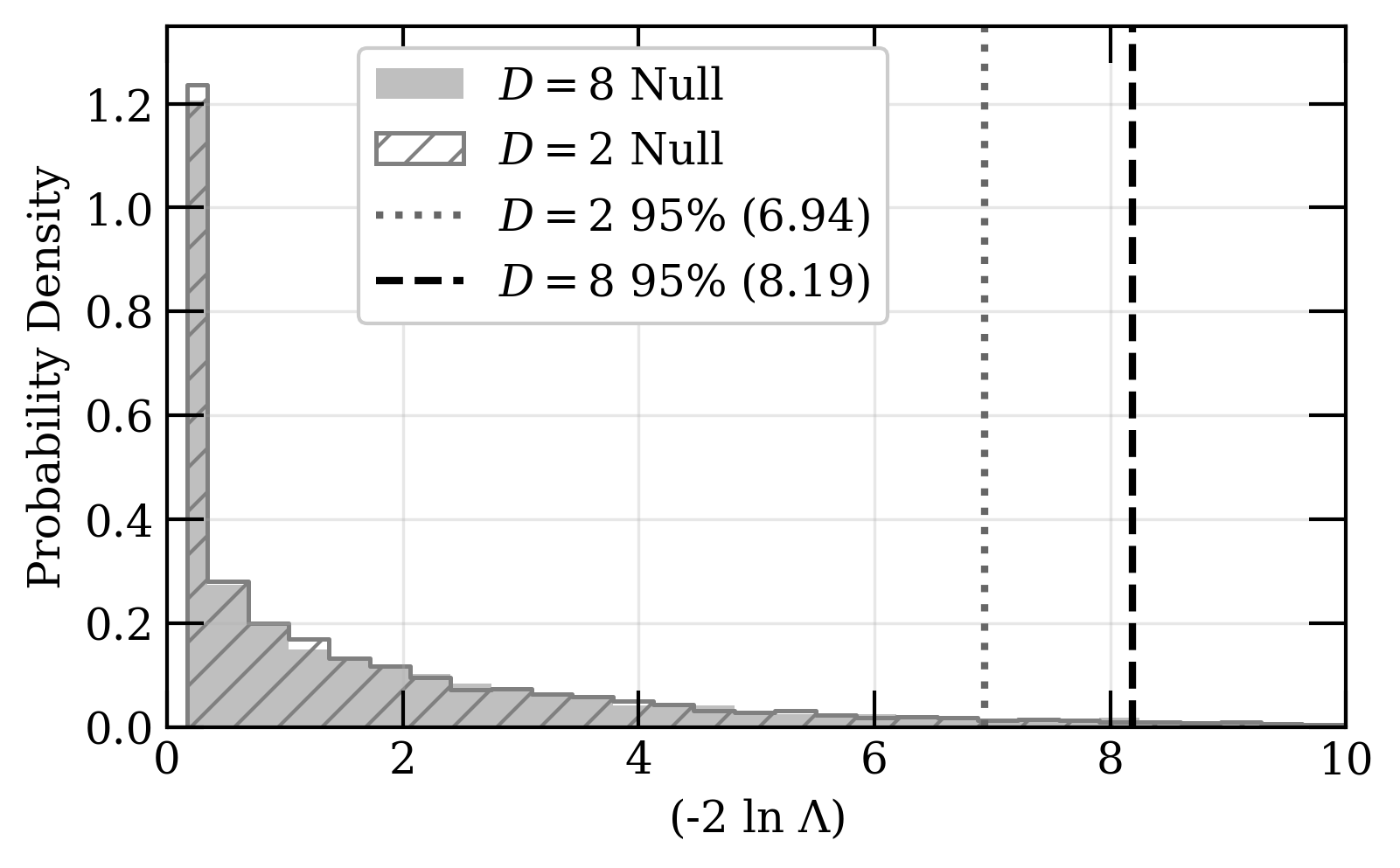} % Replace with your actual filename
    \caption{\textbf{Empirical null distributions.} Distributions of the Likelihood Ratio Test (LRT) statistic, $-2 \ln \Lambda$, under the independent branching hypothesis ($\beta_1 = 0$). The results are shown for truncation depths $D=2$ (hatched) and $D=8$ (solid gray). The vertical dashed lines indicate the empirically calibrated 95th percentile thresholds. As the depth $D$ increases, this leads to a heavier-tailed null distribution and a higher detection threshold (6.94 for $D=2$ versus 8.19 for $D=8$). This may be due to correlations between the branching dependence of related individuals contributing to the observed set. }
    \label{fig:null_dist}
\end{figure}

We simulate the null distribution for $K = 2$ offspring, $D = 2$ and $D = 8$. In each case, we evolve the tree for at least $D + 1$ generations, until its size exceeds 50 individuals after a given generation's branching event. With a sample of 50 branching events, we perform a likelihood search over an equally spaced grid of sample points, $\beta_0 \in [-1.0, 1.0]$ with $\Delta \beta_0 = 0.1$, $\theta \in [0.0, 1.0]$ with $\Delta \theta = 0.05$, and $\beta_{1} \in [0.0, 1.0]$ with $\Delta \beta_{1} = 0.05$. The maximal likelihood is determined for each of 10,000 trials. These distributions are plotted in \ref{fig:null_dist}. 

\begin{figure}[th]
    \centering
    \includegraphics[width=\textwidth]{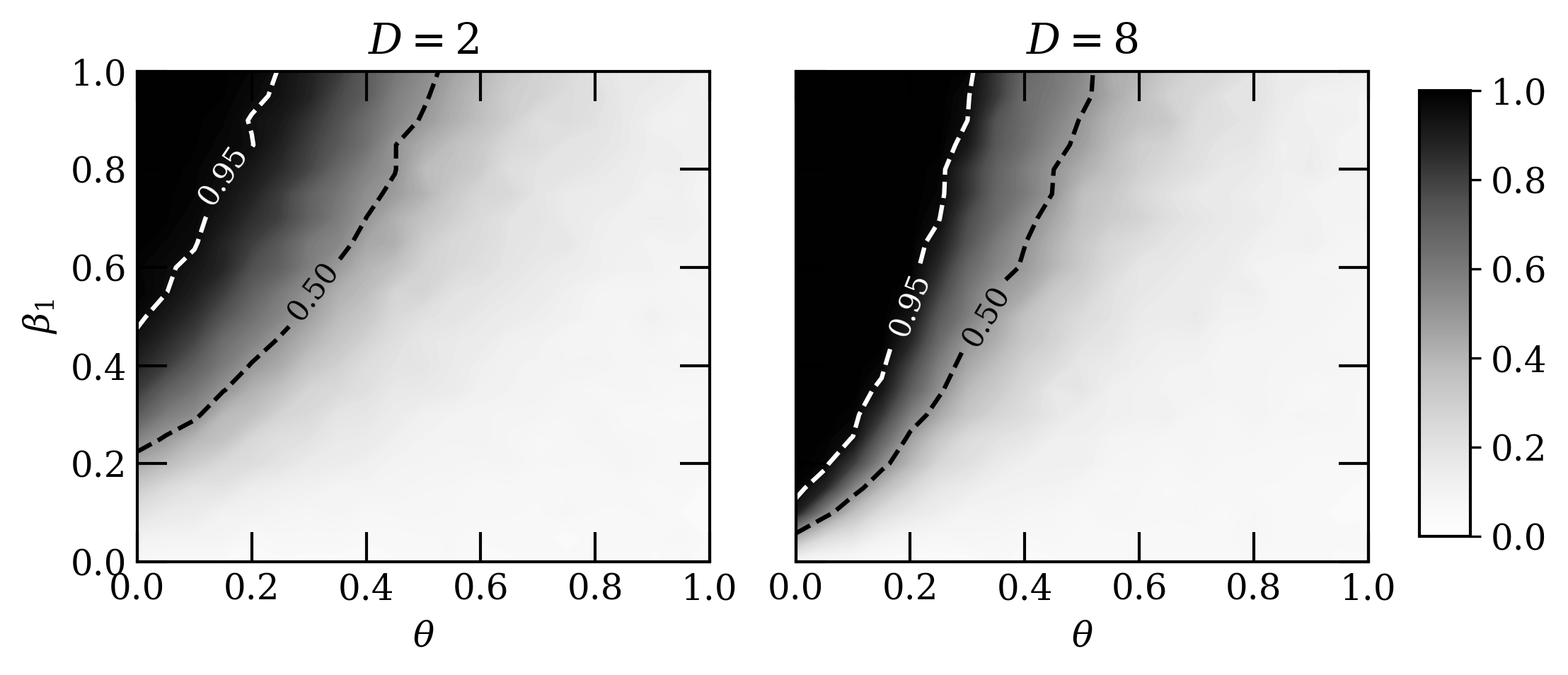} % Replace with your actual filename
    \caption{\textbf{Statistical power to detect genealogical reinforcement}. We present the statistical power as a function of the reinforcement strength $\beta_1$ and decay parameter $\theta$. The heatmaps compare truncation depths $D=2$ (left) and $D=8$ (right) for a sample size of $N=50$ and baseline $\beta_0=0$. Darker regions indicate higher power. The dashed contour lines mark the 50\% and 95\% detection thresholds. As the truncation depth increases from $D=2$ to $D=8$, the high-power contours expand significantly toward lower $\beta_1$ and higher $\theta$. This demonstrates that observing deeper genealogical relationships substantially increases statistical power, allowing for the detection of weaker or more localized reinforcement signals.}
    \label{fig:power_scan}
\end{figure}

\subsection{Characterization of Rejection Power}
To characterize the rejection power of our statistical analysis for $K = 2$, $D = 2$ and $D = 8$, we evolve the branching process under a selection of assumed model parameters, $(\beta_{0} = 0, \beta_{1}, \theta)$. We sample the same values of $\theta \in [0.0, 1.0]$ and $\beta_{1} \in [0.0, 1.0]$ as those used in our likelihood scan. For each parameter combination, we perform 1000 trials, and compare this resulting distribution of test statistics to the null distribution. We define statistical power as the probability of correctly rejecting the null hypothesis of independent branching when the alternative process is true. 

Specifically, for each simulated trial, we compute the Likelihood Ratio Test statistic, $-2 \ln \Lambda$. We then determine whether this statistic exceeds the empirically calibrated 95th percentile of the null distribution. The fraction of trials exceeding this 95th percentile is the statistical power. Maps of the statistical power for the cases of $D = 2$ and $D = 8$ are presented in Figure \ref{fig:power_scan}. Future work will investigate the impact of shared genealogical structure and correlated observations on such statistical tests.

\section{Tree-Valued Stochastic Processes}

\subsection{Tree-Valued Multitype Branching Processes}

In the case of a finite-depth genealogical dependence, branching at an individual level depends on the number or structure of relatives of relevant, bounded distance. While an individual's type could essentially be related to this bounded number of relatives, the type distributions expected for offspring are not determined based solely on the parent, and are instead determined by the tree evolution in the next generation. Given that the process evolution depends on extended genealogical information, we anticipate that the system behaves as a multitype branching process at the level of the relevant tree structure, $\mathcal{K} \in \mathcal{K}_{D}$.

It is useful to view the tree of depth, $D$, as the reproductive object. As the number of offspring is bounded, and the depth of genealogical dependence is finite, only a finite number of tree types are available, 
\begin{equation}
    \mathcal{K}_{D} = \{ \mathcal{K} : \mathcal{K} \textrm{ rooted, depth } \leq D, \textrm{ offspring } \leq 2 \}.
\end{equation}

At each generation's branching event, the distribution of new offspring per individual is determined by the tree structure. From the perspective of the process, this original tree type begets zero, one or two new trees of depth, $k_{\textrm{MRCA}}$. The number of these offspring trees and tree type is determined by the parent tree type. This is a form of multitype branching over trees \citep{athreya1972branching, georgii2003supercriticalmultitypebranchingprocesses}. 

\begin{theorem}
    Branching processes with genealogical dependence of bounded distance $2D$ are represented by tree-valued multitype branching processes over a finite-size type space corresponding to trees of depth $D$.
\end{theorem}

\begin{proof}
    To establish this equivalence, we map our branching process with genealogical dependence to a tree-valued multitype process on a finite type space, and show that the Markov property holds. 

    \textbf{Step 1: Finiteness of the type space.}
    Let $K$ be the maximal number of offspring and $2D$ the truncation distance of genealogical dependence. Individual branching depends only on the local tree structure of distance $2D$, corresponding to a tree of generational depth $D$. As both $K$ and $D$ are finite, the number of possible rooted trees of depth $D$ is finite. We denote this finite type space $\mathcal{K}_{D}$. Let $|\mathcal{K}_D|$ denote the number of distinct tree types. 

    \textbf{Step 2: The multitype representation.} 
    At any generation $n$, partition the living population into a collection of local trees of depth $D$. The state of the process is the vector of counts of tree types, $Z_n = (Z_n^{(1)}, \dots, Z_n^{(|\mathcal{K}_D|)})$, where $Z_n^{(i)}$ is the number of type-$i$ trees present. Following the classical formulation \citep{athreya1972branching}, the offspring distribution is defined by the probability generating functions, $f^{(i)}(\mathbf{s})$, for enumerated tree types, $\mathbf{s} = (s_1, \dots, s_{|\mathcal{K}_D|})$:
    \begin{equation}
        f^{(i)}(\mathbf{s}) = \sum_{j_1, \dots, j_{|\mathcal{K}_D|} \ge 0} p^{(i)}(j_1, \dots, j_{|\mathcal{K}_D|}) s_1^{j_1} \cdots s_{|\mathcal{K}_D|}^{j_{|\mathcal{K}_D|}},
    \end{equation}
    where $p^{(i)}(j_1, \dots, j_{|\mathcal{K}_D|})$ is the probability that a type-$i$ parent tree produces $j_k$ trees of type $k$. As the reproduction probability $p(v)$ of any individual $v$ depends only on its local tree, the transition rules of the entire process are entirely determined by the current count of trees. 

    \textbf{Step 3: The branching property.} 
    As the branching of each individual depends only on relatives within a finite distance of $2D$, the evolution of a tree containing a group of relatives depends only on itself. Each tree within the current state, $Z_n$, evolves independently to the next generation. 

    \textbf{Conclusion.} 
    As the transition from $Z_n$ to $Z_{n+1}$ depends only on $Z_n$, this process satisfies the Markov property. Additionally, the individual trees evolve or branch independently given the current state. The process $Z_n$ is then exactly a multitype branching process on the finite state space $\mathcal{K}_D$.
\end{proof}

Let $|\mathcal{K}^{D}|$ represent the number of distinct tree types. In our specific case of two maximal offspring and a general depth $D$, $|\mathcal{K}^{D}|$ is found from the recurrence relation,
\begin{equation}
    |\mathcal{K}^{D}| = 1 + |\mathcal{K}^{D - 1}| + \binom{ |\mathcal{K}^{D - 1}| + 1}{2} = \dfrac{(|\mathcal{K}^{D - 1}| + 1)(|\mathcal{K}^{D - 1}| + 2)}{2}.
\end{equation}

We consider the number of each tree type present at generation, $n$. Let,
\begin{equation}
    Z_{n} = \left( Z_{n}^{(1)}, \dots, Z_{n}^{|\mathcal{K}^{D}|} \right).
\end{equation}
represent the population of tree types. We introduce the offspring matrix, $\mathbf{M} = (m_{ij})$, where
\begin{equation}
    m_{ij} = \mathbb{E} \left[ \text{number of type-\(j\) trees produced by a type-\(i\) tree} \right].
\end{equation}
These expectations can be determined from the tree structure and model parameters describing dependence or reinforcement. This value, $m_{ij}$ describes the expected contribution of tree type $i$ to tree type $j$ in one generation. 

In our specific model of genealogical reinforcement, we account for our parameterization in $\beta_{0}, \beta_{1}$ and $\theta$ for a specific depth of influence, $D$. For a given type-$i$ tree, we determine the probability of 0, 1, or 2 offspring for each individual in the current generation. This allows us to construct the distribution of type-$j$ trees of depth $D$ expected from the type-$i$ tree. 

\begin{example}[Construction of the Mean Matrix for $D=2$]
Consider the case $D = 2$. Using the recurrence relation, there are $|\mathcal{K}^{2}| = 10$ distinct tree types. We will compute a row of the mean matrix $\mathbf{M}$ corresponding to a single individual with two offspring, where one branch has one additional offspring and the other has two offspring. 

We begin by determining the probability of $\xi(v)$ offspring for each individual in the current generation. We label the individuals $v_{1}$, $v_{2}$, and $v_{3}$, corresponding to the sibling pair of two individuals, and the single child. We have, 
\begin{align}
    \eta_{1,2} &= \beta_0 + \beta_1 \left(e^{-\theta} + e^{-2\theta} \right), \\
    \eta_{3} &= \beta_0 + \beta_1 \left(e^{-2\theta} \right). 
\end{align}
These log-odds correspond to reproduction probabilities, 
\begin{align}
    p(v_{1,2}) &= \dfrac{1}{1 + \exp(-\eta_{1,2})}, \\
    p(v_{3}) &= \dfrac{1}{1 + \exp(-\eta_{3})}.
\end{align}
Given these probabilities, the number of offspring for each individual follows a Binomial distribution. The probabilities that individuals $v_{1}$, $v_{2}$, and $v_{3}$ produce exactly \(k \in \{0, 1, 2\}\) offspring are:
\begin{align}
    \mathbb{P}(\xi(v_{1,2}) = k; p(v_{1,2})) &= \binom{2}{k} p(v_{1,2})^k \big(1 - p(v_{1,2})\big)^{2-k}, \\
    \mathbb{P}(\xi(v_{3}) = k; p(v_{3})) &= \binom{2}{k} p(v_{3})^k \big(1 - p(v_{3})\big)^{2-k},
\end{align}

To compute the probability of transitioning to a specific set of tree types in the next generation, we consider all possible combinations of offspring counts. Let $k_1, k_2, k_3 \in \{0, 1, 2\}$ denote the number of offspring produced by $v_1, v_2$, and $v_3$, respectively. Because their branching events are conditionally independent, the joint probability of this configuration is the product of their individual probabilities:
\begin{equation}
    \mathbb{P}(k_1, k_2, k_3) = \mathbb{P}(k_1; p(v_1)) \cdot \mathbb{P}(k_2; p(v_2)) \cdot \mathbb{P}(k_3; p(v_3)).
\end{equation}

This generational branching event introduces a new group of offspring. The evolution of this group will only depend on the previous $D$ generations. These offspring do not depend on the oldest generation of depth $D + 1$. As this single common ancestor at depth $D + 1$ had two offspring, this previous tree produces two offspring trees, where subsequent branching of either tree is uncorrelated. In general, each tree type may produce 0, 1 or 2 offspring trees. 

The distribution of expected tree types is determined from the probability of each offspring count combination. To populate the row of the mean matrix $\mathbf{M}$ corresponding to this parent tree, we sum over all $3^{3} = 27$ possible configurations. The expected number of type-$j$ trees produced is given by:
\begin{equation}
\begin{split}
    m_{i,j} &= \sum_{k_1=0}^2 \sum_{k_2=0}^2 \sum_{k_3=0}^2 \mathbb{P}(k_1, k_2, k_3) \\
    &\quad \times \big(\text{number of type-}j\text{ trees from configuration } (k_1, k_2, k_3)\big).
\end{split}
\end{equation}
This process is repeated for each of the $|\mathcal{K}^{2}| = 10$ parent tree types to construct the full $10 \times 10$ mean matrix $\mathbf{M}$. The summation accounts for identical tree types reached through distinct combinations. 

\end{example}

As $Z_n$ is a discrete-time, finite-state multitype branching process, standard results from Perron-Frobenius theory and the classical theory of multitype processes apply \citep{baake2025criticalmultitypebranchingfunctional}. The spectral radius of $\mathbf{M}$, denoted $\rho(\mathbf{M})$, determines the criticality of the process. The process is subcritical if $\rho(\mathbf{M}) < 1$, critical if $\rho(\mathbf{M}) = 1$, and supercritical if $\rho(\mathbf{M}) > 1$. The population's distribution of tree types converges to a stable distribution given by the left eigenvector of $\mathbf{M}$. This stable distribution is the direct long-time consequence of genealogical reinforcement.

Additionally, standard generating-function techniques and martingale convergence theorems can be applied to characterize extinction probabilities, moments, and Yaglom-style conditioned limit laws. The genealogical structure of this multitype population can also be studied backward in time, resulting in a structured coalescent process \citep{Schweinsberg2003CoalescentPO, 10.1214/23-AAP1934, Cordero2025StructuredCC}. In particular, the connection between density-dependent branching and genealogy is an active area of research \citep{andré2026momentsdensitydependentbranchingprocesses}.  
%We use numerical simulations to explicitly map the critical surface $\rho(\mathbf{M}) = 1$ as a function of the model parameters $(\beta_0, \beta_1, \theta)$ and the truncation depth $D$. 

\subsection{Discussion and Outlook}

In this work, we introduced the concept of genealogical dependence within branching processes. The generating function becomes directly dependent on the genealogical structure at a finite truncation depth, $D$. In this case, the process evolution is Markov and its properties are described by a multitype branching process over a finite-size type space of trees. A similar analysis could be applied to the corresponding coalescent.

A natural extension of this work would involve analysis of genealogical dependence in the infinite depth limit, allowing $D \rightarrow \infty$. As the truncation depth grows, the state space of local tree types becomes infinite in size. The process depends on the entire historical state and would be represented as an infinite-dimensional Markov process. Depending on the assumed form of genealogical dependence or reinforcement, it may be possible to introduce an approximation scheme to reduce the dimensionality of the problem. In particular, if the strength of correlation with distant relatives decays sufficiently fast, the correlated, distant genealogy could be approximated with a bounded dimensionality. Further, we might describe the local branching as embedded in a random, dynamic system, where distant, largely uncorrelated relatives play the role of the random environment.

An additional extension would involve analysis in the large-population limit. Continuous-state branching processes, superprocesses, and descriptions of the limiting stochastic dynamics subject to ecological constraints would provide intuitive, approximatory representations of the influence of these small-scale dependencies \citep{Etheridge2000AnIT, Lambert_2005, Birkner_2021}. The genealogical lines of descent for prolific individuals in such processes may be captured via backbone decompositions \citep{kyprianou2013backbonedecompositionspatiallydependent} and spine techniques \citep{harris2019many}. A key consideration is incorporating discrete, tree-valued interactions within a continuous, measure-valued process \citep{thresh}. Analysis would likely require a more general tree-valued object \citep{evans2008realtrees}. 

\section*{Appendix A: Fisher Information and Identifiability}

To rigorously establish local identifiability, we show that the Fisher Information Matrix (FIM) is positive definite at the true parameter values. Let the parameter vector be $\boldsymbol{\phi} = (\beta_0, \beta_1, \theta)^\top$. 

The observed data for an individual $v$ is the number of offspring $\xi(v)\in \{0, 1, 2\}$, distributed as $\text{Binomial}(K = 2, p(v))$, where $p(v) = \frac{1}{1 + \exp(-\eta_v)}$. The log-likelihood for a single observation is:
\begin{equation}
    \ell_v(\boldsymbol{\phi}) = \xi(v) \log p(v) + (2 - \xi(v)) \log(1 - p(v)) + \text{const}.
\end{equation}

The score function (first derivative with respect to $\boldsymbol{\phi}$) is:
\begin{equation}
    \nabla \ell_v = (\xi(v) - 2p(v)) \nabla \eta_v,
\end{equation}
where the Jacobian of the linear predictor is:
\begin{equation}
    \nabla \eta_v = \begin{pmatrix} 
    1 \\ 
    \sum_{d} \mathbf{c}_{v}(d) e^{-\theta d} \\ 
    -\beta_1 \sum_{d} d \cdot \mathbf{c}_{v}(d) e^{-\theta d} 
    \end{pmatrix}.
\end{equation}

The Fisher Information Matrix for the full tree is the sum of the individual information matrices:
\begin{equation}
\begin{aligned}
    \mathcal{I}(\boldsymbol{\phi}) &= \sum_{v} \mathbb{E} \left[ \nabla \ell_v (\nabla \ell_v)^\top \right] \\
    &= \sum_{v} \left( 2 p(v)(1 - p(v)) \right) \nabla \eta_v (\nabla \eta_v)^\top.
\end{aligned}
\end{equation}
As $2 p(v)(1 - p(v)) > 0$ for all $p(v) \in (0,1)$, the matrix $\mathcal{I}(\boldsymbol{\phi})$ is positive definite if and only if the set of Jacobian vectors $\{\nabla \eta_v\}_{v \in \mathcal{T}_n}$ spans $\mathbb{R}^3$. Assuming the tree contains at least two distinct, nonzero distances, the vectors $(1, e^{-d_1\theta}, -d_1 e^{-d_1\theta})$ and $(1, e^{-d_2\theta}, -d_2 e^{-d_2\theta})$ are linearly independent. As established in \textbf{Case (a)} and \textbf{Case (b)} of the main text, the Jacobian vectors $\{ \nabla \eta_{v} \}_{v \in \mathcal{T}_{n}}$ are linearly independent, making the Fisher Information Matrix $\mathcal{I}(\boldsymbol{\phi})$ non-singular. This guarantees local identifiability.

\begin{remark}[Concavity]

The Hessian of the log-likelihood is:
\begin{equation}
    \nabla^2 \ell_v = -2p(v)(1-p(v)) \nabla \eta_v (\nabla \eta_v)^\top + (\xi(v) - 2p(v)) \nabla^2 \eta_v.
\end{equation}
Because the second term contains the data-dependent residual $(\xi(v) - 2p(v))$, which can be positive or negative, the finite-sample Hessian is not guaranteed to be globally negative semi-definite. While the expected likelihood is uniquely maximized (global identifiability), the finite-sample log-likelihood may be non-concave. To guarantee finding the global maximum in our numerical simulations, we utilize a global likelihood search over a grid of parameter values.
\end{remark}

\section*{Acknowledgments}
The author is grateful to Emmanuel Schertzer and Fernando Cordero for insightful discussions reflected in the development of this work.

\bibliographystyle{unsrt}
\bibliography{ref.bib} 

\end{document}